\pdfoutput=1
\documentclass[12pt]{article}

\font\grande=cmr9.5 scaled \magstep4
\font\medio=cmr9.5 scaled \magstep2
\outer\def\beginsection#1\par{\medbreak\bigskip
      \message{#1}\leftline{\bf#1}\nobreak\medskip
\vskip-\parskip
      \noindent}
\usepackage{graphicx} 
\begin{document}
\bibliographystyle {unsrt}

\titlepage

\begin{flushright}
CERN-TH-2016-069
\end{flushright}

\vspace{10mm}
\begin{center}
{\grande The first observations of wide-band interferometers}\\
\vskip 10mm
{\grande and the spectra of relic gravitons}\\
\vspace{15mm}
 Massimo Giovannini 
 \footnote{Electronic address: massimo.giovannini@cern.ch} \\
\vspace{1cm}
{{\sl Department of Physics, 
Theory Division, CERN, 1211 Geneva 23, Switzerland }}\\
\vspace{0.5cm}
{{\sl INFN, Section of Milan-Bicocca, 20126 Milan, Italy}}

\vspace*{1cm}
\end{center}

\centerline{\medio  Abstract}
\vspace{5mm}
Stochastic backgrounds of relic gravitons of cosmological origin extend from frequencies of the order of the aHz 
up to the GHz range. Since the temperature and polarization anisotropies constrain the low frequency normalization of the spectra, in the concordance paradigm the strain amplitude corresponding to the frequency window of wide-band interferometers turns out to be, approximately, nine orders of magnitude smaller than 
the astounding signal recently reported and attributed to a binary black hole merger. The 
backgrounds of relic gravitons expected from the early Universe are compared with the stochastic foregrounds
stemming from the estimated multiplicity of the astrophysical sources. It is suggested that while the 
astrophysical foregrounds are likely to dominate between few Hz and 10 kHz, relic gravitons with frequencies exceeding 100 kHz represent a potentially uncontaminated signal for the next generation of high-frequency detectors currently under scrutiny.  
\vskip 0.5cm

\nonumber
\noindent

\vspace{5mm}

\vfill
\newpage
Stochastic backgrounds of relic gravitons have been envisaged as a genuine general relativistic effect well before 
the formulation of any of the conventional models aiming at a specific account 
of the early stages of the evolution of our Universe \cite{gr}. Since the governing equations of the tensor modes 
of the geometry are not invariant under the Weyl rescaling of the four-dimensional metric,
relic gravitational waves are amplified thanks to the pumping action of the space-time curvature itself.
Prior to the pioneering investigations of Ref. \cite{gr} the wave equations of the tensor modes were believed to be Weyl-invariant 
 as it happens in the case of chiral fermions and electromagnetic waves in four space-time dimensions. 
After nearly forty years of analyses and speculations, stochastic backgrounds of relic gravitons are now one of the most plausible 
predictions of general relativity and of various classes of inflationary models \cite{rub}.  In the pivotal scenario the production of relic gravitons is characterized by decreasing frequency spectra whose amplitudes and slopes are
simultaneously fixed (at the conventional pivot wavenumber $k_{p} = 0.002\, \mathrm{Mpc}^{-1}$)  by $r_{T}=  {\mathcal A}_{T}/{\mathcal A}_{{\mathcal R}}$ where ${\mathcal A}_{T}$ and ${\mathcal A}_{{\mathcal R}}$ denote, respectively,  the amplitudes\footnote{In the concordance paradigm (often dubbed $\Lambda$CDM paradigm where $\Lambda$ denotes the dark energy component and CDM stands for the cold dark matter contribution) the spectral slope $n_{T}$ and the slow-roll parameter $\epsilon$  are expressible in terms of $r_{T}$ according to the so-called consistency relations stipulating that $n_{T} = - 2\epsilon = - r_{T}/8$ \cite{WMAP9,PLANCK}. } of the tensor and of the scalar power spectra.  
The lowest frequency range of the graviton spectra is therefore ${\mathcal O}(\mathrm{few})$ aHz ($1\, \mathrm{aHz} = 10^{-18} \, \mathrm{Hz}$) 
and it corresponds to the pivot frequency $\nu_{p} = k_{p}/(2\pi) = 3.092\,\,\mathrm{aHz}$. The maximal frequency range depends instead on the post-inflationary 
transition; in the sudden approximation, customarily adopted in the concordance lore, the highest frequency of the spectrum is a fraction of the GHz ($1\,\mathrm{GHz} = 10^{9}\,\,\mathrm{Hz}$):
\begin{eqnarray}
\nu_{max}  &=& \frac{1}{4\pi} \biggl(2 \pi \, r_{T} \, {\mathcal A}_{{\mathcal R}} \, \Omega_{\mathrm{R}0} \biggr)^{1/4}\, \sqrt{H_{0} \, M_{P}}
\nonumber\\
&=& 0.3 \,\biggl(\frac{r_{T}}{0.1}\biggr)^{1/4} 
\biggl(\frac{{\mathcal A}_{\mathcal R}}{2.41 \times 10^{-9}}\biggr)^{1/4}
\biggl(\frac{h_{0}^2 \Omega_{\mathrm{R}0}}{4.15 \times 10^{-5}}\biggr)^{1/4} \,\mathrm{GHz},
\label{nineth}
\end{eqnarray}
where $\Omega_{\mathrm{R}0}$ is the fraction of critical energy density attributed to massless particles (photons and neutrinos in the vanilla $\Lambda$CDM paradigm \cite{WMAP9,PLANCK})
while $h_{0}$ is the present value of the Hubble rate $H_{0}$ in units of $10^{2} \mathrm{km}/(\mathrm{sec}\,\times\mathrm{Mpc})$.  Equation (\ref{nineth}) is derived 
by redshifting the frequency from the end of the inflationary epoch to the present time. At the end of inflation the maximal frequency of the 
spectrum coincides approximately with the Hubble rate during inflation. Today this frequency is clearly different and it depends, 
in principle, on the whole post-inflationary history.  The simplest way to estimate this quantity is to consider the case where the reheating 
occurs suddenly: in this case the end of inflation coincides with the onset of the radiation dominated phase.  Recalling that, in the sudden reheating approximation, the total redshift between the end of inflation and the present time is given by $2 \sqrt{H_{0} \, M_{P}} [2 \,\Omega_{\mathrm{R}0}/(\pi r_{T} {\mathcal A}_{{\mathcal R}})]^{1/4}$ 
the result of Eq. (\ref{nineth}) follows immediately. This observation suggests that the early variations of the space-time curvature can therefore be assessed, with a fair degree of confidence, by scrutinizing the spectra of the relic gravitons at low but especially at high frequencies. 

Whenever the (transverse and traceless) tensor modes of the four-dimensional geometry 
evolve in a homogeneous and isotropic background of Friedmann-Robertson-Walker type,
 each of the two tensor polarizations follows the action of a minimally coupled scalar field so that, ultimately,  
their energy density is\footnote{In the framework of the $\Lambda$CDM paradigm \cite{WMAP9,PLANCK}, we shall assume a conformally flat background 
geometry $\overline{g}_{\mu\nu} = a^2(\tau) \eta_{\mu\nu}$ where $a(\tau)$ is the scale factor and $\eta_{\mu\nu}$ is the Minkowski metric 
with signature mostly minus; the tensor fluctuation of the geometry is defined as $\delta_{t} g_{ij} = - a^2 \, h_{ij}$.}: 
\begin{equation}
{\mathcal T}_{0}^{0} = \frac{1}{8 \ell_{P}^2 a^2 } \biggl[ \partial_{\tau} h_{ij} \partial_{\tau} h_{ij} + \partial_{k} h_{ij} \partial_{k}h_{ij} \biggr],\qquad h_{i}^{i} = \partial_{i}h^{i}_{j} =0, \qquad \ell_{P} = \sqrt{8 \pi G},
\label{first}
\end{equation}
where $\tau$ is the conformal time coordinate and $\ell_{P}$ is the Planck length. While different prescriptions can be used to assign the energy-momentum pseudo-tensor of the relic gravitons, they all coincide when the corresponding wavelengths are shorter than the Hubble radius (see last paper of \cite{BR} for a derivation of various pseudo-tensors and for their mutual comparison in different regimes). The result of Eq. (\ref{first}) follows from the analysis if Ford and Parker (see second paper of Ref. \cite{BR}) but the Landau-Lifshitz approach leads to the same results in the regime of short wavelengths, as we shall briefly discuss later on. 

The energy density of Eq. (\ref{first}) also determines the tensor power spectrum ${\mathcal P}_{T}(k, \tau)$ measuring the amplitude the two-point function at equal times
in Fourier space\footnote{Note that $p_{ij}(\hat{k}) = (\delta_{i j} - \hat{k}_{i} \hat{k}_{j})$ is the transverse projector.}:
\begin{equation}
\langle h_{ij}(\vec{k},\tau) \, h_{mn}(\vec{p},\tau) \rangle = \frac{2\pi^2}{k^3} {\mathcal P}_{T}(k,\tau) \, {\mathcal S}_{ijmn}(\hat{k}) \delta^{(3)}(\vec{k} +\vec{p}),
\label{second}
\end{equation}
where ${\mathcal S}_{ijmn}(\hat{k}) = [p_{m i}(\hat{k}) p_{n j}(\hat{k}) + p_{m j}(\hat{k}) p_{n i}(\hat{k}) - p_{i j}(\hat{k}) p_{m n}(\hat{k})]/4$. 
The two-point function for the conformal time derivative of the amplitudes (i.e. $\langle \partial_{\tau}  h_{ij}\,\partial_{\tau} h_{mn}\rangle$)  is given by an expression formally analog to Eq. (\ref{second}) but characterized by a different power spectrum  denoted hereunder by ${\mathcal Q}_{T}$. 
With these specifications the energy density of the relic gravitons is simply $\rho_{gw} = \langle {\mathcal T}_{0}^{0} \rangle$,
where the average is taken over the quantum state minimizing the Hamiltonian of the relic gravitons. 
Along a  complementary perspective the expectation value appearing in Eq. (\ref{second}) can be viewed as an ensemble average over classical amplitudes with respect to a suitable stochastic process. In units of the critical energy density $\rho_{crit}$, the relation between the tensor power spectrum and the energy density per logarithmic\footnote{The natural logarithms will be denoted by ``$\ln$'' while the common logarithms will be denoted by ``$\log$''.}  interval of comoving wavenumber (i.e. the spectral energy density) is given by:
\begin{eqnarray}
\Omega_{gw}(k,\tau) &=& \frac{1}{\rho_{crit}} \frac{d \rho_{gw}}{d \ln{k}} = \frac{1}{24 H^2 a^2} 
\biggl[k^2{\mathcal P}_{T}(k, \tau) + {\mathcal Q}_{T}(k,\tau)\biggr] 
\nonumber\\
&\to& \frac{k^2}{12 H^2 a^2} {\mathcal P}_{\mathrm{T}}(k,\tau)\biggl[ 1 + {\mathcal O}\biggl(\frac{{\mathcal H}^2}{k^2}\biggr)\biggr], \qquad k \tau \gg 1.
\label{third}
\end{eqnarray}
The final result of Eq. (\ref{third}) holds when the modes are inside the Hubble radius since, in this case, ${\mathcal Q}_{T}\to[ k^2 {\mathcal P}_{T}(k,\tau) + {\mathcal O}({\mathcal H}^2)]$. In the opposite limit (i.e. $k/{\mathcal H} \ll 1$) we have instead that ${\mathcal Q}_{T}(k,\tau) \to [{\mathcal H}^2 {\mathcal P}_{T}(k,\tau) + {\mathcal O}(k^2)]$ where, as usual, ${\mathcal H} = a H$. 
The energy densities (and pressures) derived within different approaches coincide, to leading order\footnote{In the case of the Ford-Parker energy-momentum pseudo-tensor \cite{BR} the numerical factor in front of the correction ${\mathcal O}({\mathcal H}^2/k^2)$  in Eq. (\ref{third}) is $1/2$.  Using the Landau-Lifshitz approach the correction appearing in Eq. (\ref{third}) is then given by $-7/2$ (instead of $1/2$). 
}, when the corresponding  wavelengths are inside the Hubble radius, i.e. $k > {\mathcal H}$.   

The tensor power spectrum, the spectral amplitude $S_{h}(\nu,\tau)$ (measured in units of $\mathrm{Hz}^{-1}$)  and the dimensionless strain amplitude 
are all related: since ${\mathcal P}_{T}(2 \pi \nu, \tau)= 4 \nu S_{h}(\nu,\tau)$, Eq. (\ref{third}) also implies 
\begin{equation}
S_{h}(\nu,\tau) = \frac{3 {\mathcal H}^2}{4\pi^2 \nu^3} \Omega_{gw}(\nu,\tau)  \to  7.981\times 10^{-43} \,\,\biggl(\frac{100\,\mathrm{Hz}}{\nu}\biggr)^3 \,\, h_{0}^2 \,\Omega_{gw}(\nu,\tau_{0})\,\, \mathrm{Hz}^{-1},
\label{fourth}
\end{equation}
where, as already remarked,  $\nu=k/(2\pi)$ denotes the comoving frequency in natural units. Finally the dimensionless strain amplitude 
obeys $h_{c}^2(\nu,\tau_{0}) = \nu S_{h}(\nu,\tau_{0})$ so that  $h_{c}(\nu,\tau_{0})$ becomes explicitly:
\begin{equation}
h_{c}(\nu,\tau_{0}) = 8.933 \times 10^{-21} \biggl(\frac{100 \,\, \mathrm{Hz}}{\nu}\biggr) \, \sqrt{h_{0}^2 \,\Omega_{gw}(\nu, \tau_{0})}.
\label{fifth}
\end{equation} 
The conventional inflationary models followed by a sudden reheating  imply, in the case $r_{T}=0.1$ \cite{WMAP9,PLANCK}, that $h_{0}^2 \Omega_{gw}(\nu,\tau_{0}) = {\mathcal O}(10^{-16.8})$ \cite{TR,mg3}; we are considering here 
frequencies ${\mathcal O}(100\, \mathrm{Hz})$ and we assume that the tensor spectral index does not run (see in this respect, the last paper of \cite{mg2}). For the same frequency range\footnote{The operating window of wide-band interferometers is between few Hz and roughly $10$ kHz. At low frequencies the seismic noise dominates while at intermediate and high frequencies the thermal and shot noises dominate. Barring for further suppressions of the signal to noise ratio due to the overlap reduction function (accounting for the relative location of the two correlated interferometers) the maximal sensitivity occurs for $\nu = {\mathcal O}(100) \,\,\mathrm{Hz}$ and this will be the fiducial frequency 
adopted throughout this discussion (see e.g. first three papers of Ref. \cite{mg3} and references therein).}  (compatible with the sensitivity of wide-band interferometers) Eq. (\ref{fifth}) gives $h_{c} = {\mathcal O}(10^{-29})$: while this estimate 
consistently assumes $r_{T} = {\mathcal O}(0.1)$, future determinations of $r_{T}$ might even determine an absolute normalization of the 
spectra at the pivot frequency $\nu_{p}$.

The recent detection of gravitational waves reported by the LIGO/Virgo Collaboration \cite{LIGO} 
corresponds to a dimensionless strain amplitude $h_{c} = {\mathcal O}(10^{-21})$ which is between eight and nine 
orders of magnitude larger than the stochastic background produced by the conventional inflationary models predicting, from Eq. (\ref{fifth}),
$h_{c} = {\mathcal O}(10^{-29})$. The observed burst of gravitational radiation (attributed to a merger of black holes) 
does not have an electromagnetic counterpart: both the Swift and the Fermi LAT(Large Area Telescope) satellites searched, without success, 
in the optical, ultraviolet, $x$-rays and $\gamma$-rays \cite{SWIFT}. This strategy could be useful for the near future 
insofar as only  with the direct observation of a specific electromagnetic counterpart to the gravitational signal it will be plausible 
to infer that gravitational waves travel at the speed of light (at least within a cocoon corresponding to the distance of the source). 
Absent direct confirmations of the LIGO/Virgo results,  there are reasonable hopes  that many more bursts of similar kind will be eventually observed 
with the forthcoming observational campaigns. Depending on the estimates, the rates of black hole  mergers range from ${\mathcal O}(50) \, \mathrm{Gpc}^{-3}\, \mathrm{yr}^{-1}$ 
to ${\mathcal O}(300) \, \mathrm{Gpc}^{-3}\, \mathrm{yr}^{-1}$. If this is the case we can not only expect to have many more signals 
but also a potentially new stochastic foreground coming from many unresolved sources of gravitational radiation. The typical 
amplitude of this hypothetical (and not yet detected) gravitational wave foreground will lead to\footnote{We assume the Planck determination 
of $h_{0}$ and the expectations published in \cite{LIGO} for $\Omega_{gw}$, namely $\Omega_{gw} = 1.1^{2.7}_{-0.9} \times 10^{-9} [\nu/(25 \, \mathrm{Hz})]^{2/3}$
for $25 \mathrm{Hz} < \nu \leq 100 \mathrm{Hz}$. For $\nu > 100 $ Hz the signal is exponentially suppressed (see, in particular, the last paper of Ref. \cite{LIGO}).}
\begin{equation}
h_{0}^2 \, \Omega^{(for)}_{gw} = 4.9\times 10^{-10} \,\biggl(\frac{h_{0}}{0.67}\biggr)^2 \,\biggl(\frac{\nu}{25 \, \mathrm{Hz}}\biggr)^{2/3}, \qquad 25\,\, \mathrm{Hz} < \nu \leq 100\,\,\mathrm{Hz}.
\label{sixth0}
\end{equation}
The stochastic foreground of Eq. (\ref{sixth0}) can potentially mask the stochastic background of relic gravitons coming from the early Universe. 
Indeed, we shall now show that the relic graviton backgrounds of primordial origin 
will always be smaller than the plausible expectations Eq. (\ref{sixth0}). However we shall also 
demonstrate that for frequencies exceeding $10$ kHz the preceding conclusion can be evaded.

To scrutinize the high-frequency behavior of the spectra and before the explicit numerical results, it is useful to consider the following approximate 
parametrization for the relic component\footnote{Note that $R_{\nu}$ denotes the fraction of neutrinos in the radiation plasma i.e. 
$R_{\nu} = r_{\nu}/(r_{\nu} +1)$ where $r_{\nu} = \rho_{\nu}/\rho_{\gamma}= 0.681 (N_{\nu}/3)$ and $N_{\nu}$ is the number of massless neutrino families.}:
\begin{eqnarray}
h_{0}^2 \,\Omega_{gw}(\nu,\tau_{0}) &=& {\mathcal N}_{\rho}(r_{T}, \, \Omega_{\mathrm{R}0}) \,T^2_{low}(\nu/\nu_{\mathrm{eq}}, R_{\nu})\, T^{2}_{high}(\nu/\nu_{s}, \alpha) \biggl(\frac{\nu}{\nu_{\mathrm{p}}}\biggr)^{n_{\mathrm{T}}} e^{- 2 \,\beta\,\nu/\nu_{max}}, 
\label{seventh0}\\
{\mathcal N}_{\rho}(r_{T}\, \Omega_{\mathrm{R}0}) &=& 4.165 \times 10^{-15}\,\, r_{T}\,\, \biggl(\frac{h_{0}^2 \Omega_{\mathrm{R}0}}{4.15\times 10^{-5}}\biggr),
\label{seventh}
\end{eqnarray}
where the parameter $\beta = {\mathcal O}(1)$ depends upon the width of the transition between the inflationary phase and the subsequent radiation dominated phase;
for different widths of the post-inflationary transition we can estimate $0.5 \leq \beta \leq 6.3$ \cite{mg3}. The maximal frequency $\nu_{max}$ appearing in Eq. (\ref{seventh}) has been already introduced in Eq. (\ref{nineth}) while $\nu_{\mathrm{eq}}$ is defined as:
\begin{equation}
\nu_{eq} =  1.317 \times 10^{-17} \biggl(\frac{h_{0}^2 \Omega_{\mathrm{M}0}}{0.1364}\biggr) \biggl(\frac{h_{0}^2 \Omega_{\mathrm{R}0}}{4.15 \times 10^{-5}}\biggr)^{-1/2}\,\, \mathrm{Hz},
\label{eight}
\end{equation}
where $\Omega_{\mathrm{M}0}$ denotes the present critical fraction of dusty matter in the $\Lambda$CDM paradigm.

In Eq. (\ref{seventh}) $T_{low}$ and $T_{high}$ denote, respectively,  the transfer functions at low and high frequencies. To transfer the spectrum inside the Hubble radius the procedure is to integrate numerically the equations of the tensor modes. This procedure for the derivation of $T_{low}$ and $T_{high}$ 
has been discussed in detail in the last paper of Ref. \cite{TR}. Approximations to the full transfer function (valid in the case 
 of spectra produced during a stiff phase and in a waterfall transition)  have been derived, respectively  in Ref. \cite{mg2} (see in particular second and third papers) 
 and in the fourth paper of Ref. \cite{mg3}. In the simple limit $R_{\nu} \to 0$  the transfer function across equality is given by \cite{TR}:
\begin{equation}
T_{low}(\nu/\nu_{\mathrm{eq}},0) = \sqrt{1 + c_{1}\biggl(\frac{\nu_{\mathrm{eq}}}{\nu}\biggr) + b_{1}\biggl(\frac{\nu_{\mathrm{eq}}}{\nu}\biggr)^2},\qquad c_{1}= 0.5238,\qquad
b_{1}=0.3537.
\label{ten}
\end{equation}
Similarly, denoting with $\nu_{s}$ the frequency at which the approximate scale invariance of the spectral energy density is broken we will have that $T^2_{high} \to 1$ 
for $\nu \ll \nu_{s}$ while  $T^2_{high} \to (\nu/\nu_{s})^{\alpha}$ for $\nu \gg \nu_{s}$. Potential violations 
of approximate scale-invariance for frequencies larger than a putative frequency $\nu_{s}$ arise in various situations 
including a prolonged stiff post-inflationary phase, a delayed reheating or even the presence of spectator fields triggering waterfall transitions \cite{mg2}. In what follows 
$\nu_{s}$ will be referred to as the frequency of the ankle since it defines the beginning of the high-frequency 
branch where the spectral energy density can be sharply increasing. According to Eq. (\ref{ten}), $T_{low} \to 1$ for $\nu \gg \nu_{eq}$ but the realistic situations further suppressions are expected. 
At least two effects are not captured by Eq. (\ref{seventh}): the free-streaming of collisionless species and the dark-energy transition. The neutrino free-streaming produces an effective anisotropic stress leading ultimately to an integro-differential equation (see, for instance, \cite{wnu}).  Assuming that the only collisionless 
species in the thermal history of the Universe are the neutrinos (which are massless in the concordance paradigm \cite{WMAP9,PLANCK}), the amount 
of suppression of $h_{0}^2 \, \Omega_{gw}$  for  $ \nu_{eq} < \nu < \nu_{bbn}$ can be parametrized by the function ${\mathcal F}(R_{\nu}) = 1 -0.539 R_{\nu} + 0.134 R_{\nu}^2$ where $\nu_{bbn}$ is the comoving frequency that corresponding to the Hubble rate at nucleosynthesis:
\begin{equation}
\nu_{bbn}= 
2.252\times 10^{-11} \biggl(\frac{g_{\rho}}{10.75}\biggr)^{1/4} \biggl(\frac{T_{bbn}}{\,\,\mathrm{MeV}}\biggr) 
\biggl(\frac{h_{0}^2 \Omega_{\mathrm{R}0}}{4.15 \times 10^{-5}}\biggr)^{1/4}\,\,\mathrm{Hz} \simeq 0.01\, \mathrm{nHz},
\label{ten0}
\end{equation}
where  $g_{\rho}$ denotes the effective number of relativistic degrees of freedom entering the total energy density of the plasma and $T_{bbn}$ is the temperature of big-bang nucleosynthesis.  The  shallow suppression of $h_{0}^2 \Omega_{gw}$ in the range $ \nu_{eq} < \nu < \nu_{bbn}$ is then proportional 
to ${\mathcal F}^2(0.405)= 0. 645$ (for $N_{\nu} = 3$ and $R_{\nu} = 0.405$). The second effect not included in Eq. (\ref{nineth}) is the damping effect associated 
with the (present) dominance of the dark energy component\footnote{The redshift of $\Lambda$-dominance is given by 
$1 + z_{\Lambda} = (\Omega_{de}/\Omega_{M0})^{1/3}$; in the $\Lambda$CDM paradigm $\Omega_{\mathrm{de}} \equiv \Omega_{\Lambda}$
and the damping across $z_{\Lambda}$ reduces $h_{0}^2\, \Omega_{gw}$ by a factor $(\Omega_{\mathrm{M}0}/\Omega_{\Lambda})^2= {\mathcal O}(0.2)$. }. This effect is comparable with the suppression due to 
the neutrino free streaming\footnote{There is a third effect reducing the quasi-flat plateau implied by Eq. (\ref{ten}) in the limit $\nu \gg \nu_{eq}$: the variation of the effective number of relativistic species \cite{wnu,mg3}. In the case of the minimal standard model this would imply that the reduction will be ${\mathcal O}(0.38)$.} (see e.g. the last paper of Ref. \cite{mg3}).  

Thanks to the preceding semi-analytical arguments we can therefore infer that any departure 
from approximate scale-invariance beyond $\nu_{s} ={\mathcal O}(\mathrm{nHz})$ will lead to a signal exceeding astrophysical foregrounds but only for sufficiently large frequencies, i.e. above ${\mathcal O}(100\, \mathrm{kHz})$. 
 \begin{figure}[!ht]
\centering
\includegraphics[height=6.9cm]{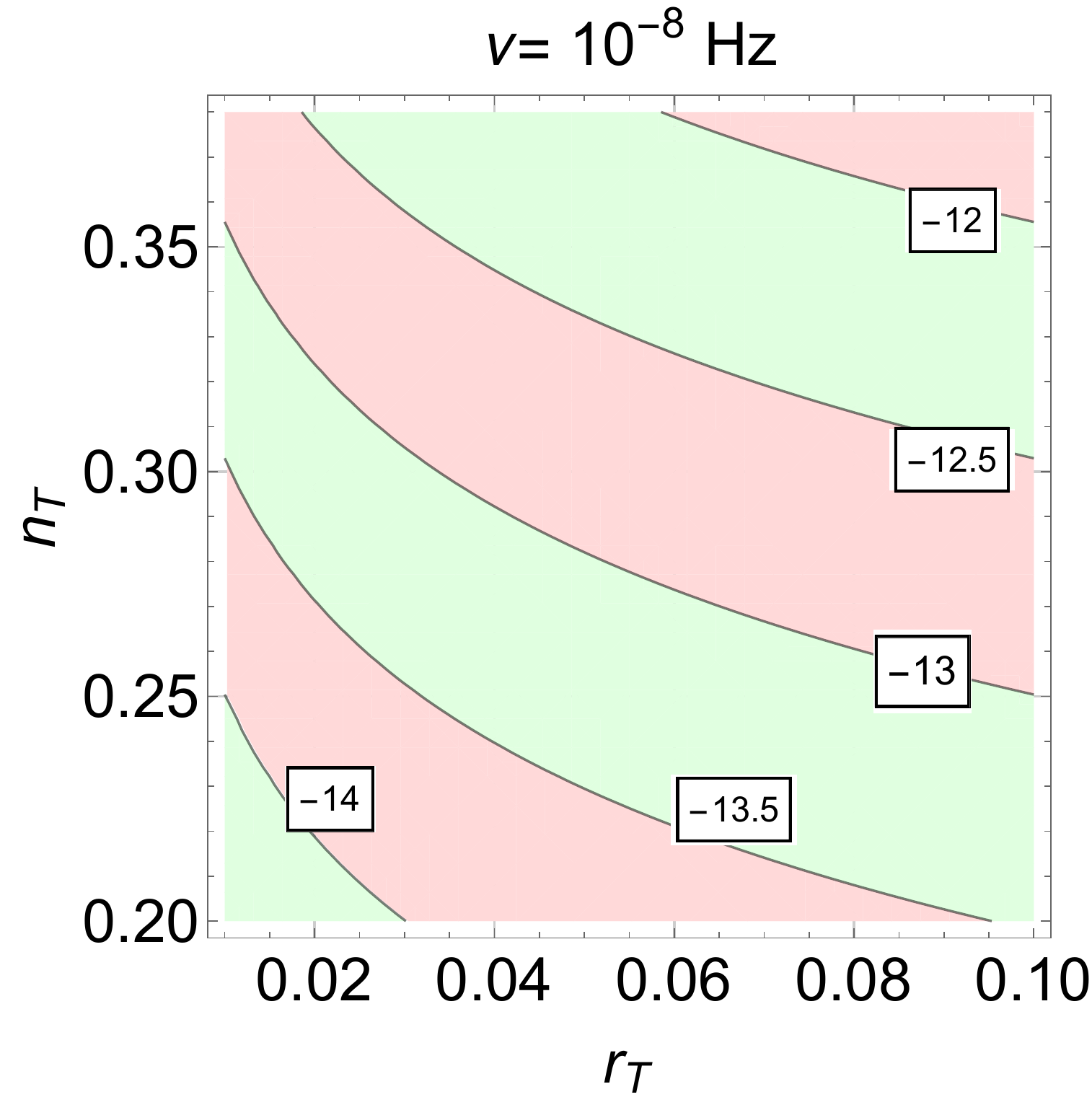}
\includegraphics[height=6.9cm]{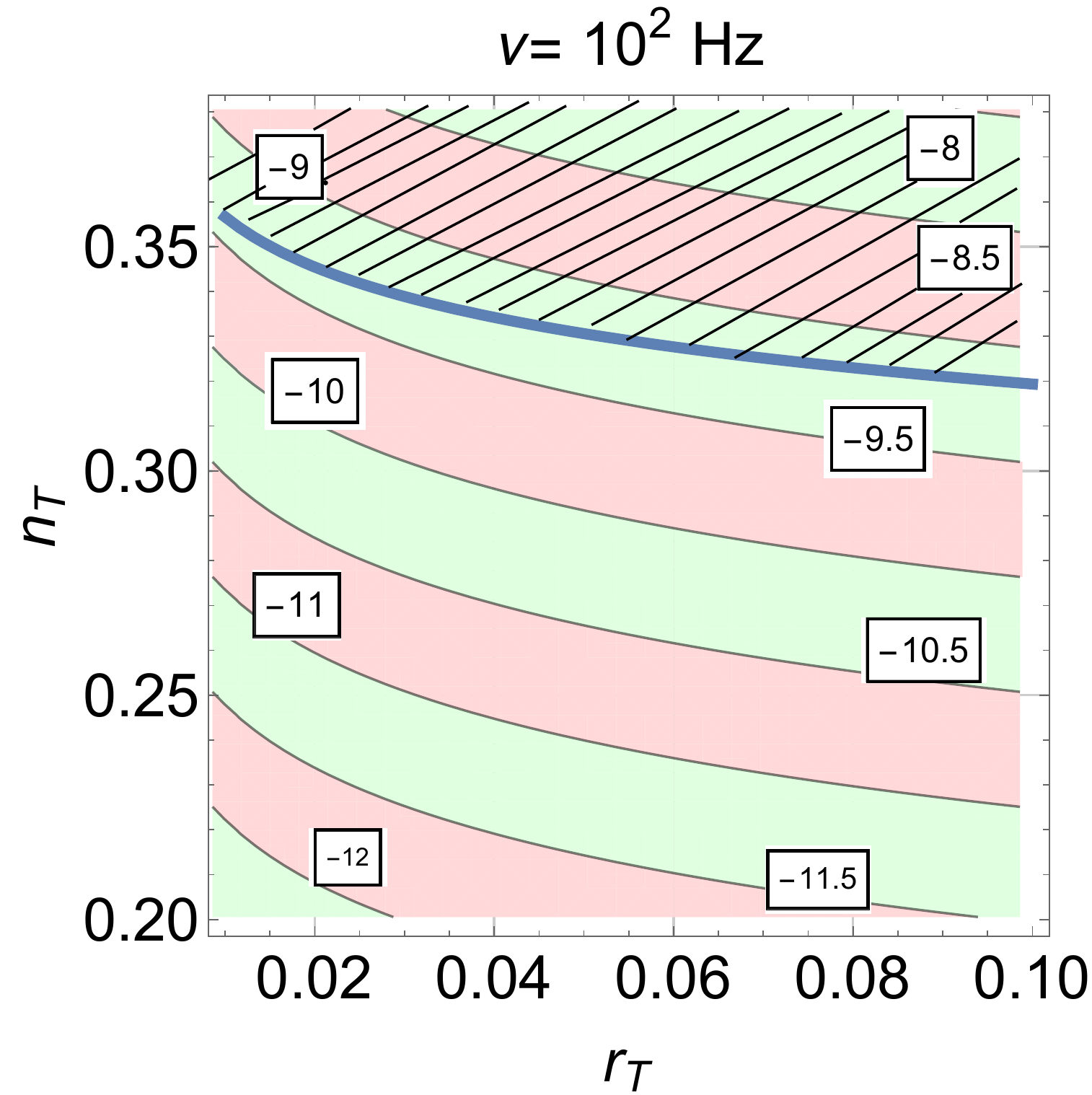}
\caption[a]{In both plots the contours correspond to a different value of the common logarithm of $h_{0}^2 \Omega_{gw}$;  $r_{T}$ and $n_{T}$ are independently assigned so that the consistency relation are violated. In the plot on the left $\nu$ corresponds 
to the pulsar timing frequency. In the plot on the right the big bang nucleosynthesis constraint at the LIGO/Virgo frequency is illustrated. }
\label{Figure1}      
\end{figure}
 A large stochastic background for frequencies ${\mathcal O} (100\, \mathrm{Hz})$ may correspond to the situation where 
$T_{high} \to 1$ but $n_{T}$ and $r_{T}$ are {\em independently assigned}. In this case
the consistency relations are violated and we can see from Fig. \ref{Figure1} that the allowed region corresponds to $n_{T} < {\mathcal O}(0.37)$, assuming $r_{T}\leq 0.1$. 
\begin{figure}[!ht]
\centering
\includegraphics[height=6.9cm]{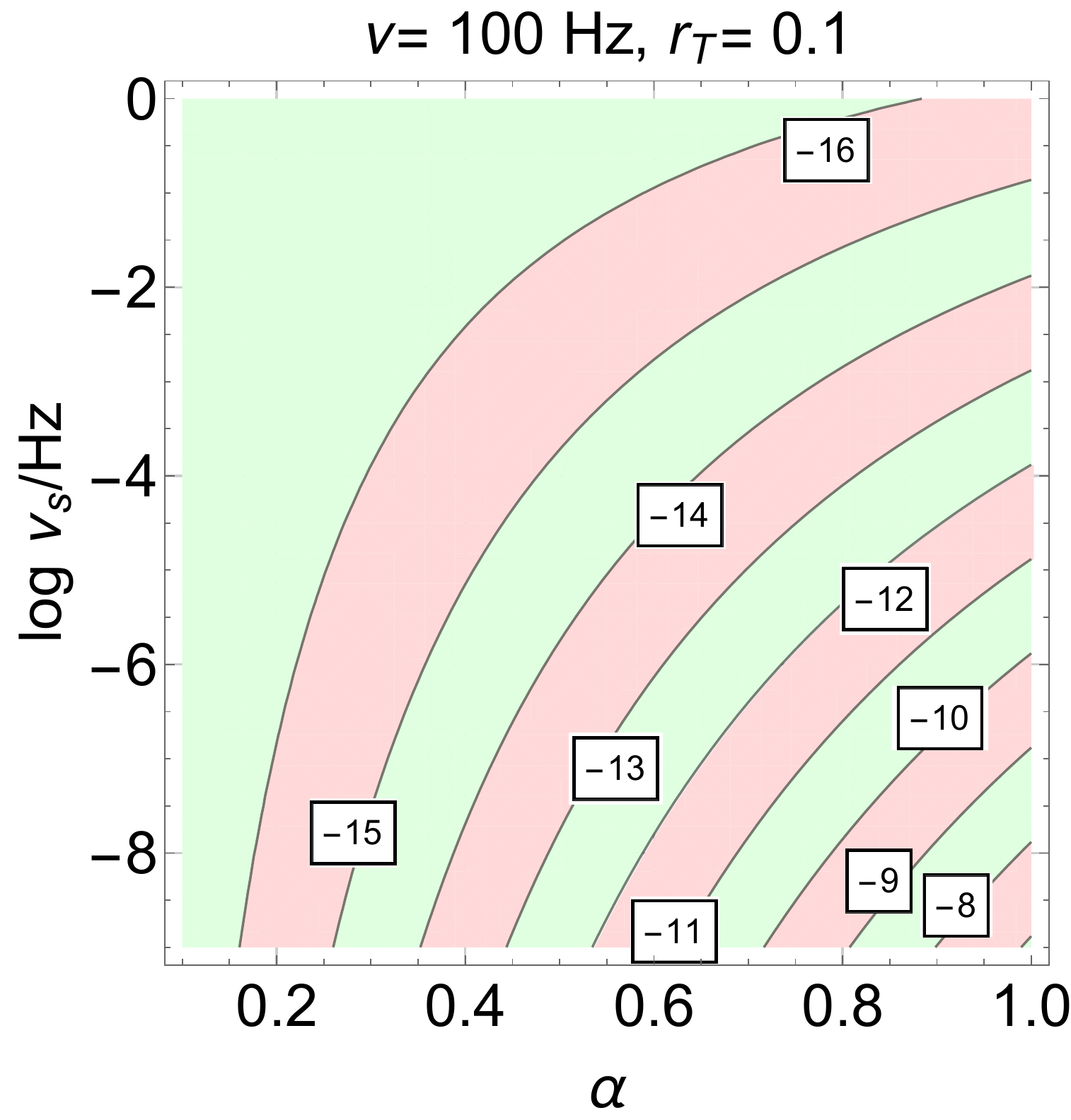}
\includegraphics[height=6.9cm]{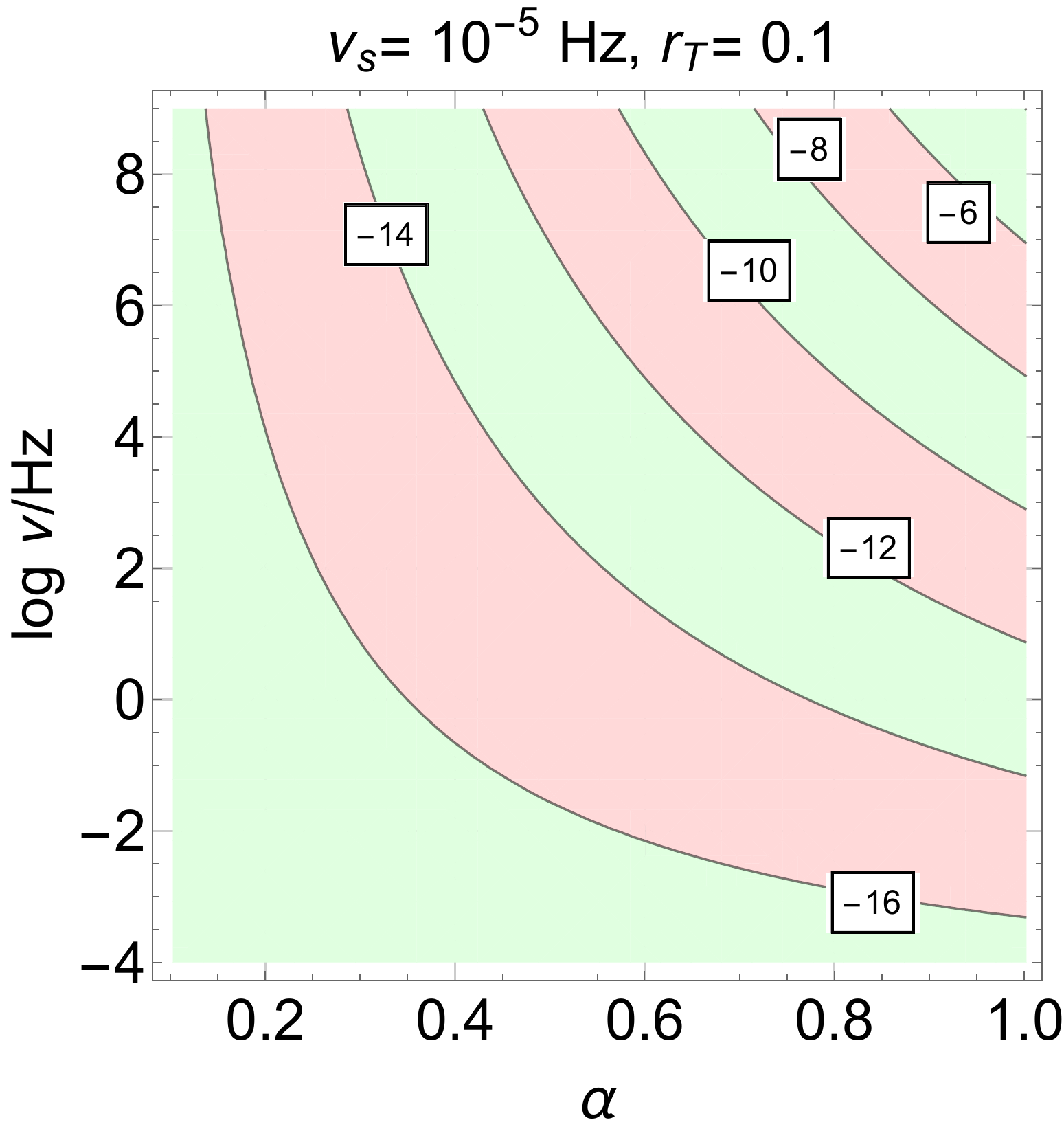}
\includegraphics[height=6.9cm]{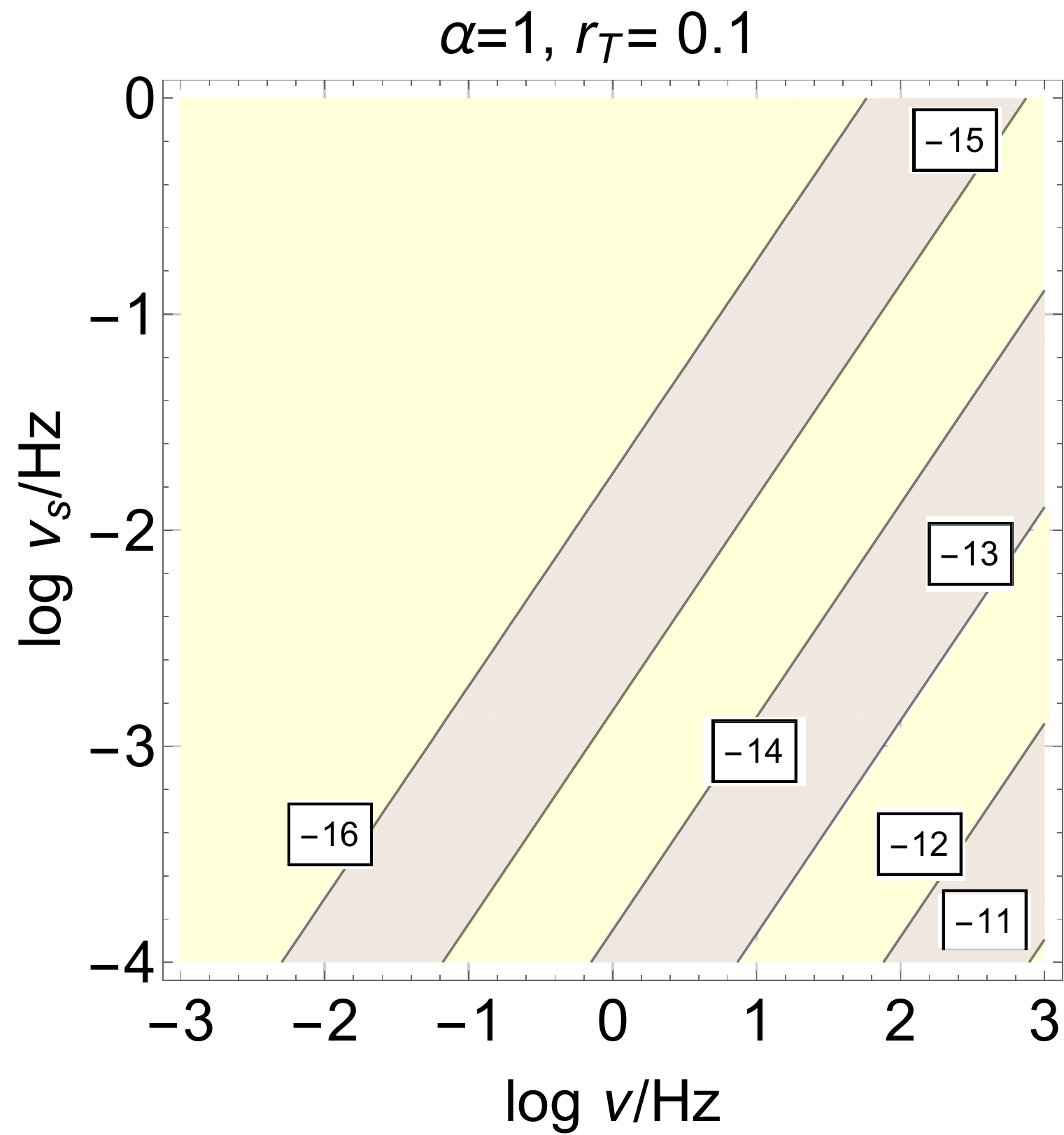}
\includegraphics[height=6.9cm]{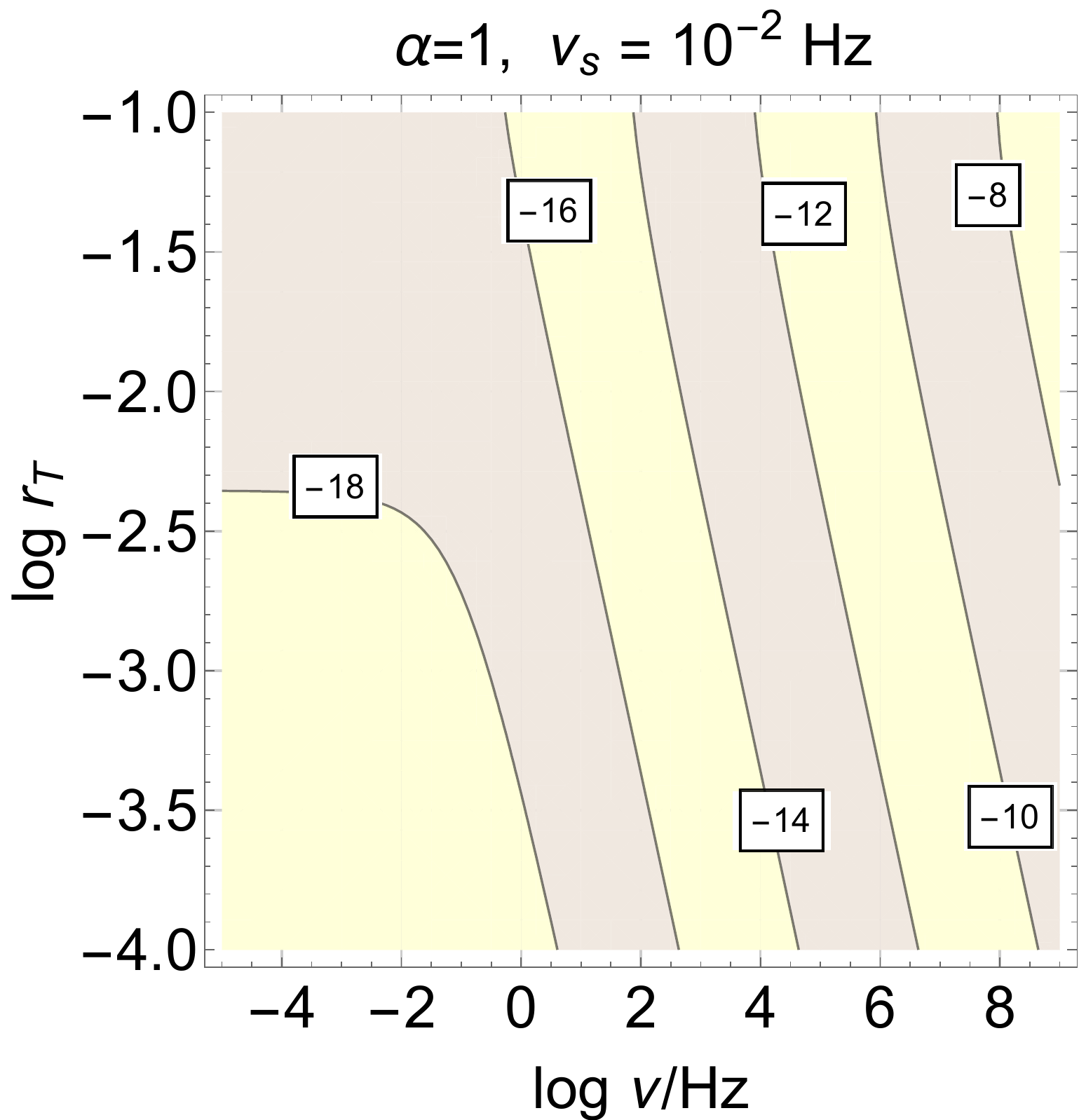}
\caption[a]{The same contours of Fig. \ref{Figure1} are illustrated in terms of $\log{h_{0}^2 \Omega_{gw}}$  for the four-dimensional parameter space 
consisting of the tensor to scalar ratio  $r_{T}$, the spectral frequency $\nu$, the frequency of the ankle $\nu_{s}$ and 
the high-frequency slope $\alpha$. We have assumed the consistency relations and the other fiducial determinations \cite{WMAP9,PLANCK} of the remaining parameters of the $\Lambda$CDM scenario supplemented by tensors. }
\label{Figure2}      
\end{figure}
In the plot 
on the left of Fig. \ref{Figure1} the pulsar timing measurements impose $\Omega_{gw}(\nu_{pulsar},\tau_{0}) < 1.9\times10^{-8}$
which is an upper bound at a typical frequency $\nu_{pulsar} \simeq \,10^{-8}\,\mathrm{Hz}$, roughly corresponding to the inverse of the observation time along which the pulsars timing has been monitored \cite{PULSAR}.  The Parkes 
 pulsar timing array  brings the limit down to $2.3\times 10^{-10}$. 
While the plot on the left of Fig. \ref{Figure1} demonstrates that for $0.1 \leq n_{T} \leq 0.4$ the pulsar limits are safely satisfied,
the plot on the right  shows clearly that $h_{0}^2 \Omega_{gw}$ can be sufficiently large for $\nu = 100$ Hz 
but the resulting signal is of the same order of the presumed astrophysical foreground. Indeed, using $\nu = 100$ Hz, Eq. (\ref{sixth0})
 implies that the stochastic foreground can be as large as $h_{0}^2 \Omega_{gw}^{(for)} = 10^{-9.3}$: this value is barely compatible with 
 the big bang nucleosynthesis bound illustrated in Fig. \ref{Figure1} with the shaded area. 
We remind that the big-bang nucleosynthesis limit sets an indirect constraint  
on the extra-relativistic species (and, among others, on the relic gravitons) at the time when light nuclei have been firstly formed \cite{BBN}. For historical reasons 
this constraint is often expressed in terms of $\Delta N_{\nu}$ representing the contribution of supplementary 
neutrino species but the extra-relativistic species do not need to be fermionic. If the additional species are 
relic gravitons we have: 
\begin{equation}
h_{0}^2  \int_{\nu_{bbn}}^{\nu_{\mathrm{max}}}
  \Omega_{gw}(\nu,\tau_{0}) d\ln{\nu} = 5.61 \times 10^{-6} \Delta N_{\nu} 
  \biggl(\frac{h_{0}^2 \Omega_{\gamma0}}{2.47 \times 10^{-5}}\biggr).
\label{BBN1}
\end{equation}
The bounds on $\Delta N_{\nu}$ range from $\Delta N_{\nu} \leq 0.2$ 
to $\Delta N_{\nu} \leq 1$ implying  
that the integrated spectral density  is between $10^{-6}$ and $10^{-5}$. In the $(r_{T}, \, n_{T})$ plane of Fig. \ref{Figure1} the 
various labels on the curves denote the common logarithm of $h_{0}^2 \, \Omega_{gw}$ and the shaded area of the plot on the 
left is forbidden by the bound of Eq. (\ref{BBN1}). Since  the relic graviton spectrum extends beyond $10$ kHz with an 
increasing spectrum (i.e. $n_{T} >0$) the integral of Eq. (\ref{BBN1}) is dominated by the highest 
frequency so that $h_{0}^2\Omega_{gw}$ must be even smaller around $100$ Hz.  
 
The example of Fig. \ref{Figure1} is simple enough but unconventional: 
 the positivity of the spectral index (i.e. $n_{T} >0$) would demand a model based either on a non-minimal gravity theory or on 
a violation of the dominant energy condition (see, in this respect, the third paper of \cite{mg2}). 
The same features of  Fig. \ref{Figure1} hold  also when  the the consistency relations are not violated but 
the spectral energy density is not (approximately) scale-invariant in the high-frequency limit, as it happens in the case 
of Eq. (\ref{seventh0}).  In Fig. \ref{Figure2} we illustrate the contours of constant $\log{h_{0}^2 \Omega_{gw}}$ for different choices of the four parameters involved in the 
discussion namely, $\nu$, $\nu_{s}$, $r_{T}$ and $\alpha$. The values of $n_{T}$ and $r_{T}$ are both constrained by the temperature and polarization anisotropies thanks to the consistency relations. 
The values of $\alpha$ and $\nu_{s}$ will instead be considered as free parameters with some plausible physical limitations. In case the inflationary phase is followed by an epoch where 
the plasma is dominated by a stiff source $\alpha \leq 1$ (up to logarithmic corrections); in the same context $\nu_{s} = {\mathcal O}(\mathrm{mHz})$ (see first 
and second papers of \cite{mg2}). Waterfall fields can also produce steep spectra with $\alpha > 1$ but with $\nu_{s} > \mathrm{kHz}$ (see fourth paper of Ref. \cite{mg3}).
Since, at least in principle, the signal must be relevant for the LIGO/Virgo operating window, the latter case is less interesting than the former.  This is why, in Fig. \ref{Figure2}, we just 
reported the results for $\alpha \leq 1$.  The spectral energy density in critical units 
always undershoots the value of the foreground of Eq. (\ref{sixth0}) for $\nu = {\mathcal O}(100)$ Hz (top left plot in Fig. \ref{Figure2}). The regions where, apparently, the signal is large enough are excluded 
by the nucleosynthesis bound of Eq. (\ref{BBN1}). This happens in particular in the case of the plot on the top left corner of Fig. \ref{Figure2} 
where the curves leading to $h_{0}^2 \Omega_{gw} = {\mathcal O}(10^{-8})$ and ${\mathcal O}(10^{-6})$ are in fact excluded. The pulsar bound, in this context, does not play any role since $\nu_{s} \gg 10^{-8}$ Hz (see the bottom right plot of Fig. \ref{Figure2}). The approximate choice of parameters $\alpha \to 1$ and $\nu_{s} = {\mathcal O}(10^{-2}) \mathrm{Hz}$ (top and bottom right plots in Fig. \ref{Figure2}) always produces the largest signal.

There are some who might worry about the possibility that above the kHz many more sources could show up. This is difficult to foresee since 
the high-frequency detectors are not yet in an advanced stage of development (see following paragraph). From the theoretical 
viewpoint we would like to mention three interesting possibilities \cite{speculations} and contrast them with the present findings. 
In the first paper of Ref. \cite{speculations} the authors describe an interesting idea to detect gravity waves at high frequencies 
with particular attention to the frequency range of $50-300$ kHz.  As far as the possibility of a signal is concerned the authors only mention the effect of QCD axion in astrophysical stellar mass black holes. This signal occurs, according to the paper, for a typical frequency of $290$ kHz (twice 
the mass of the axion) for a Peccei-Quinn scale $10^{16}$ GeV. The signal is coherent and monochromatic 
thus completely different from the stochastic backgrounds discussed in this paper. In the second reference of \cite{speculations}
 the authors discuss potential signals of gravitational radiation (for frequencies larger than $100$ Hz) arising in a model where 
two branes are connected by a black string. Finally in the third and fourth papers of Ref. \cite{speculations} the authors suggest that, maybe, magnetized plasmas could produce intense gravitational waves. The frequencies appearing in this last paper of Ref. \cite{speculations} range between $100$ Hz and $100$ kHz with rather uncertain amplitudes.

While the recent detection of gravitational radiation cannot be independently confirmed,  
the multiplicity of the observed events is bound to increase dramatically in the near future. If this is the case, the presence of an astrophysical foreground of stochastically distributed amplitudes of gravitational waves seems to be unavoidable. These foregrounds are likely to mask the stochastic backgrounds of relic gravitons bearing the mark of the early variation of the Hubble expansion rate. For frequencies larger than the mHz the primeval spectra can be decreasing, quasi-flat or even increasing (with spiky shapes) mostly depending on post-inflationary evolution. While the astrophysical foregrounds will always be dominant at least between $25$  and $100$ Hz, over higher frequencies this conclusion can be evaded. The present analysis suggests that future plannings of space-borne interferometers  (in the mHz range) of terrestrial networks (between few Hz and $10$ kHz) should be usefully complemented by high-frequency instruments  (such as microwave resonator or wave guide detectors \cite{HF}) operating above $10$ kHz. The forthcoming generations of high-frequency detectors might be the sole hope of achieving a direct detection of cosmic backgrounds of relic gravitons free from foreseeable foreground contaminations.

\end{document}